\documentclass[twocolumn,prl,showpacs,superscriptaddress]{revtex4}
\usepackage{graphicx}
\usepackage{amsmath}
\usepackage{amssymb}
\begin{document}
\title{Exchange and collective behavior of magnetic impurities in a disordered helical metal}
\author{H\'ector Ochoa}
\affiliation{Donostia International Physics Center, 20080 San Sebasti\'an, Spain\\
and Fundaci\'on IMDEA Nanociencia, Cantoblanco 28049, Madrid, Spain}
\date{\today}
\begin{abstract}
We study the exchange interaction and the subsequent collective behavior of magnetic impurities embedded in a disordered two-dimensional (2D) helical metal. The exchange coupling follows a statistical distribution whose moments are calculated to the lowest order in $\left(p_F\ell\right)^{-1}$, where $p_F$ is the Fermi momentum of itinerant electrons and $\ell$ is the mean free path. We find that i) the first moment of the distribution decays exponentially, and ii) the variance of the interaction is long-range, however, it becomes independent of the orientation of the localized magnetic moments due to the locking between spin and momentum of the electrons that mediate the interaction. As consequence, long-range magnetic order tends to be suppressed, and a spin glass phase emerges. The formalism is applied to the surface states of a three-dimensional (3D) topological insulator. The lack of a net magnetic moment in the glassy phase and the full randomization of spin polarization at distances larger than $\ell$ excludes a spectral gap for surface states. Hence, non-magnetic disorder may explain the dispersion in results for photoemission experiments in magnetically-doped topological insulators.
\end{abstract}
\maketitle

A topological insulator is a system that supports metallic edge/surface states within the bulk gap, whose existence and integrity are protected by time-reversal symmetry \cite{rev}. In the case of a 3D topological insulator, the surface states disperse as Dirac quasi-particles; the minimal Hamiltonian consists in a Bychkov-Rashba spin-orbit coupling of the form ($\hbar=1$)
\begin{align}
\mathcal{H}_{BR}=v_F\left(\boldsymbol{\sigma}\times\mathbf{p}\right)_z,
\label{eq:H_TI}
\end{align}
where $v_F$ is the Fermi velocity of surface electrons and $\boldsymbol{\sigma}=\left(\sigma_x,\sigma_y\right)$ is a vector of Pauli matrices acting on electron spin. The inclusion of a mass term in the Hamiltonian, $M\sigma_z$, reflects the breakdown of time-reversal symmetry. This may arise as a Zeeman term due to a weak magnetic field \cite{Qu_etal} or the proximity exchange coupling to a magnetic thin film \cite{magnetic_films}. In this scenario, the system becomes a 2D quantum Hall liquid, a perfect platform for several magneto-electric effects \cite{magneto-electric}.

\begin{figure}
\begin{centering}
\includegraphics[width=1\columnwidth]{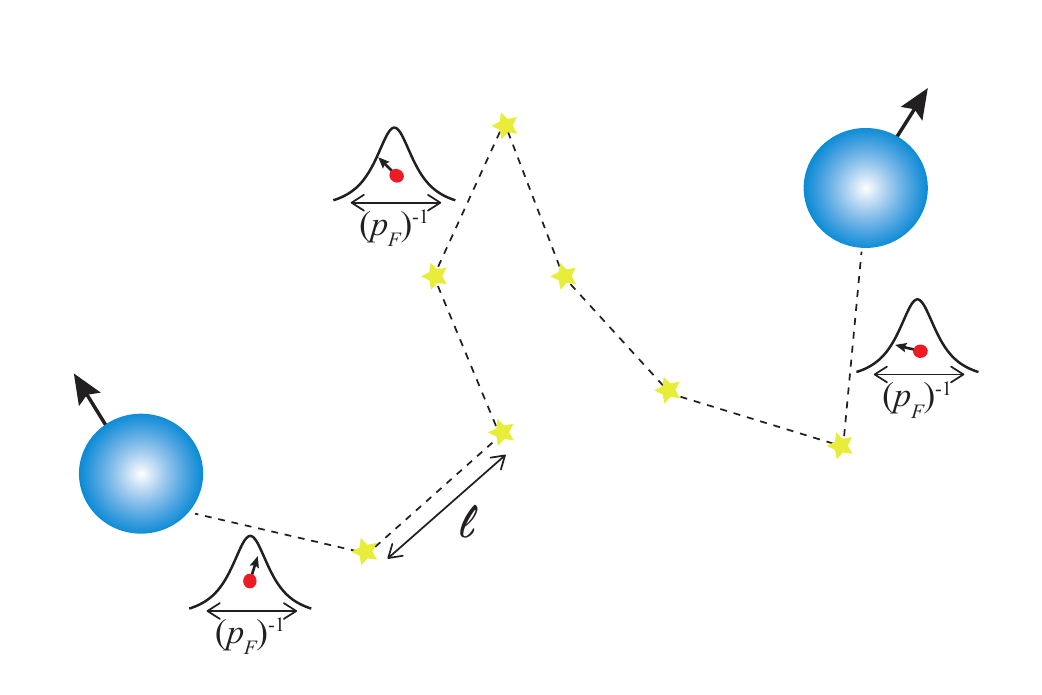}
\par\end{centering}
\caption{\label{fig:figure} Sketch of the RKKY interaction mediated by itinerant electrons in a disordered helical metal.}
\end{figure}

An interesting possibility is the deposition of magnetic adatoms, in such a way that, for high enough concentrations, $n_m$, and at low enough temperatures, $T$, a spontaneous ordering opens a gap in the spectrum, being the mass $M$ proportional to the net magnetic moment in the out-of-plane direction $\hat{z}$ \cite{Zhang1,Zhang2,Abanin_Pesin}. Experimentally, this possibility does not seem to be completely clear. Angle-resolved photoemission (ARPES) and scanning tunnel spectroscopy (STS), together with X-ray magnetic circular dichroism (XMCD) and magneto-transport experiments generate results that are in agreement with a gap opening in some cases \cite{general_yes} and gapless spectrum in others \cite{general_no} for similar experimental conditions. In the particular case of Fe adatoms deposited directly on the surface of Bi$_2$Se$_3$, for example, a gap of the order of 100 meV at $T\sim 15$ K was reported in an ARPES experiment \cite{nphys1838}, whereas later on, another ARPES experiment in similar conditions of Fe concentration and even lower temperatures reported no gap opening \cite{PRL108}, further confirmed by STS and XMCD experiments \cite{honolka2012}. This dispersion in experimental results motivates the present study about the impact of disorder in the collective behavior of magnetic adatoms placed on the surface of a 3D topological insulator.

We consider first the problem of the exchange interaction between localized spins associated to magnetic impurities embedded in a weakly disordered helical metal. The Hamiltonian reads in general
\begin{align*}
\mathcal{H}=\mathcal{H}_{it}+\mathcal{H}_{loc},
\end{align*}
where the first terms refers only to the itinerant electrons and the second term describes the coupling with localized spins,
\begin{align*}
\mathcal{H}_{loc}=J_{z}\sum_{i\in\text{m}}\text{s}_z\left(\mathbf{R}_i\right)\cdot\text{S}_i^{z}+J_{\parallel}\sum_{i\in\text{m}}\mathbf{s}\left(\mathbf{R}_i\right)\cdot\mathbf{S}_i,
\end{align*}
where the sum runs over the magnetic impurities. Here $\text{S}_i^z$, $\mathbf{S}_i=\left(\text{S}_i^x,\text{S}_i^y\right)$ are the spin operators of the impurity and $\text{s}_z\left(\mathbf{R}_i\right)$, $\mathbf{s}\left(\mathbf{R}_i\right)=\left(\text{s}_x\left(\mathbf{R}_i\right),\text{s}_y\left(\mathbf{R}_i\right)\right)$ are the spin density operators of the metal evaluated at the magnetic impurity sites,
\begin{align*}
\text{s}_{\alpha}\left(\mathbf{R}_i\right)=\sum_{\mathbf{r}}\delta^{(2)}\left(\mathbf{R}_i-\mathbf{r}\right) \sigma_{\alpha}\left(\mathbf{r}\right).
\end{align*}
The exchange is taken to be anisotropic due to the 2D nature and strong spin-orbit coupling of the system. In-plane isotropy is assumed, $J_x=J_y\equiv J_{\parallel}$.

The single-particle Hamiltonian for itinerant electrons can be written as $\mathcal{H}_{it}=\mathcal{H}_{BR}+V\left(\mathbf{r}\right)$, where the first term corresponds to Eq.~\eqref{eq:H_TI}, and the second describes the effect of disorder, $V\left(\mathbf{r}\right)=\sum_{i\in\text{nm}}\mathcal{V}_{imp}\left(\mathbf{r}-\mathbf{R}_i\right)$. Here $\mathcal{V}_{imp}\left(\mathbf{r}\right)$ is the potential created by a non-magnetic impurity or any source of disorder that preserves time-reversal symmetry. Since we do not posses a detailed expression for $\mathcal{V}_{imp}\left(\mathbf{r}\right)$ and the distribution of impurities change from sample to sample, we employ a statistical description in terms of different disorder realizations forming an ensemble of macroscopically identical replicas of the system. Assuming that the typical decay length of the impurity potential is smaller than the mean separation between scattering centers, we consider a Gaussian distribution of disorder configurations, determined by the mean free path $\ell$, or equivalently, the scattering time $\tau=\ell/v_F$, and characterized by correlators of the form 
\begin{align*}
\left\langle V\left(\mathbf{r}_1\right)V\left(\mathbf{r}_2\right)\right\rangle_{dis}=\frac{1}{2\pi\gamma\tau}\delta^{(2)}\left(\mathbf{r}_1-\mathbf{r}_2\right),
\end{align*}
where $\gamma$ is the density of states at the Fermi level, $\gamma=p_F/\left(2\pi v_F\right)$. The itinerant electrons mediate the Ruderman-Kittel-Kasuya-Yosida (RKKY) interaction \cite{RKKY} between the localized magnetic moments. For a dilute concentration of magnetic impurities this can be studied within second order perturbation theory. Then, the effective exchange coupling is determined by the static spin susceptibility of the helical electron gas. In the presence of non-magnetic disorder, this follows a statistical distribution whose moments can be computed by standard diagrammatic techniques. The first moment of the distribution is defined as $\chi_{\alpha\beta}\left(\mathbf{R}=\mathbf{R}_j-\mathbf{R}_i\right)\equiv\left\langle\chi_{\alpha\beta}\left(\mathbf{R}_i,\mathbf{R}_j\right)\right\rangle_{dis}$, where we emphasize with the notation that translation invariance is restored on average. To the lowest order in $\left(p_F\ell\right)^{-1}$ we can write
\begin{align*}
\chi_{\alpha\beta}\left(\mathbf{R}\right)=-\frac{1}{\beta}\sum_{i\omega}\mbox{Tr}\left[\sigma_{\alpha}\hat{G}\left(i\omega,\mathbf{R}\right)\sigma_{\beta}\hat{G}\left(i\omega,-\mathbf{R}\right)\right],
\end{align*}
where $\hat{G}\left(i\omega,\mathbf{R}\right)\equiv\left\langle\hat{G}\left(i\omega,\mathbf{R}_i,\mathbf{R}_j\right)\right\rangle_{dis}$ is the disorder-averaged Green operator in the Matsubara frequencies domain, and the trace is taken in the spin indices. Here we are neglecting vertex corrections, which is strictly valid at $R\equiv\left|\mathbf{R}\right|\gg\ell$, whereas the self-energy corrections are introduced by means of $\tau$ in the self-consistent Born approximation. The hierarchy of length scales of the problem for which this equation strictly applies is $p_F^{-1}\ll\ell\ll R$, as illustrated in Fig.~\ref{fig:figure}. This corresponds to a \textit{classical diffusive regime}: helical electrons connecting a pair of localized spins describe a classical diffusion path suffering several collisions with non-magnetic scatterers. In the $T\rightarrow 0$ limit, the disorder-averaged RKKY interaction reads
\begin{widetext}
\begin{align}
\label{eq:1st_moment}
\left\langle\text{H}_{\text{RKKY}}\right\rangle_{dis}=-\frac{p_Fe^{-R/\ell}}{4\pi^2v_F R^2}\sum_{i,j\in \text{m}}\left[J_z^2\sin\left(2Rp_F\right)\text{S}_i^z\text{S}_j^z+J_{\parallel}^2\sin\left(2Rp_F\right)\left(\hat{\mathbf{R}}\cdot\mathbf{S}_i\right)\cdot\left(\hat{\mathbf{R}}\cdot\mathbf{S}_j\right)-J_zJ_{\parallel}\cos\left(2Rp_F\right)\left(\mathbf{S}_i\times\mathbf{S}_j\right)_{\hat{\mathbf{z}}\times\hat{\mathbf{R}}}\right].
\end{align}
\end{widetext}
$J_{z}$ controls the exchange between the out-of-plane spin components, and $J_{\parallel}$ between the projections along the direction linking the two magnetic impurities. The last term is a Dzyaloshinskii-Moriya coupling. In the diffusive regime, the three components decay exponentially as in a conventional 2D electron gas \cite{deGennes}. This must not be interpreted as an exponential suppression of the RKKY interaction, but as the result of the randomization of its characteristic oscillatory tail \cite{rusos}. Consequently, higher moments of the distribution must be studied.

\begin{figure}
\begin{centering}
\includegraphics[width=1\columnwidth]{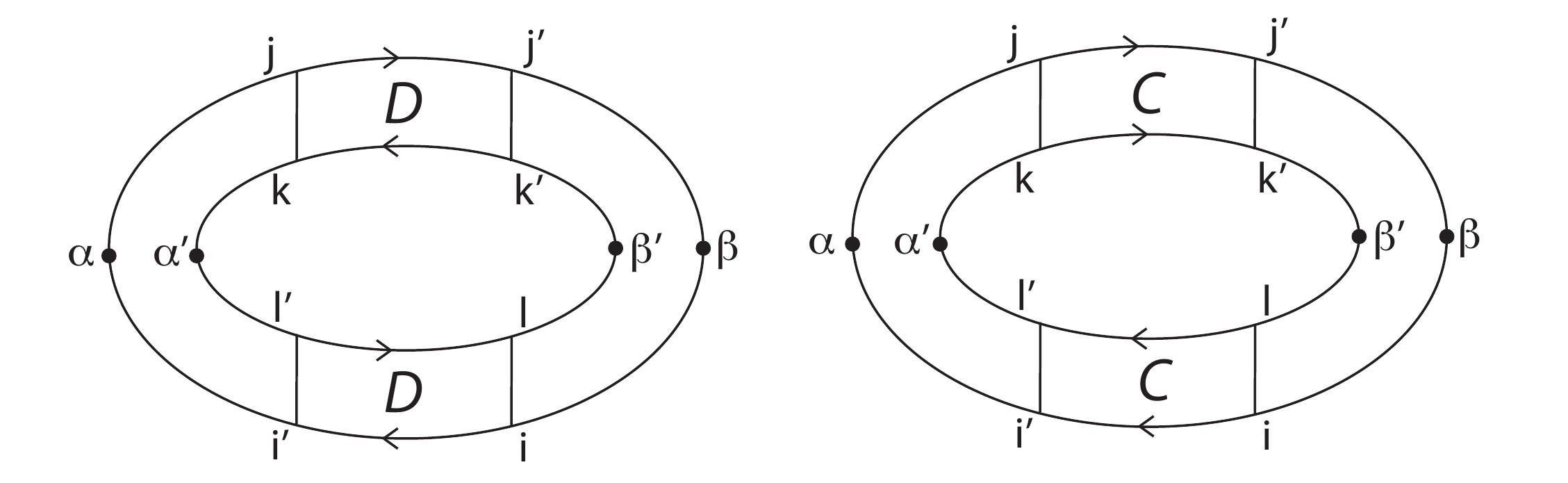}
\par\end{centering}
\caption{\label{fig:diagrams} Diffuson and cooperon ladder contributions to $\chi_{\alpha\beta}\chi_{\alpha'\beta'}$. The latin indices refer to spin components of the diffuson/cooperon kernel, and the greek indices label the spin operators at the vertex.}
\end{figure}

The calculation of the second moment of the distribution, $\chi_{\alpha\beta}\chi_{\alpha'\beta'}\left(\mathbf{R}=\mathbf{R}_j-\mathbf{R}_i\right)\equiv\left\langle\chi_{\alpha\beta}\left(\mathbf{R}_i,\mathbf{R}_j\right)\chi_{\alpha'\beta'}\left(\mathbf{R}_i,\mathbf{R}_j\right)\right\rangle_{dis}$,
is analogous to the calculation of the universal conductance fluctuations in a disordered conductor \cite{Lee_Stone}. To the lowest order in $\left(p_F\ell\right)^{-1}$, the second moment is given by both \textit{diffuson} and \textit{cooperon} ladder contributions, see Fig.~\ref{fig:diagrams}. In the absence of time-reversal symmetry breaking perturbations (excluding the localized spins) both contributions are equal. We can write
\begin{gather*}
\chi_{\alpha\beta}\chi_{\alpha'\beta'}\left(\mathbf{R}\right)=\frac{\left(2\pi\gamma\tau\right)^2}{2\beta^2}\sum_{i\omega_1,i\omega_2}\sum_{\mu,\nu,\mu',\nu'}D_{\mu\nu}\left(i\omega_1-i\omega_2,\mathbf{R}\right) 
\nonumber\\
\times D_{\mu'\nu'}\left(i\omega_1-i\omega_2,-\mathbf{R}\right)\mbox{Tr}\left[\sigma_{\alpha}\sigma_{\mu}\sigma_{\alpha'}\sigma_{\nu'}\right]
\mbox{Tr}\left[\sigma_{\beta}\sigma_{\mu'}\sigma_{\beta'}\sigma_{\nu}\right],
\end{gather*}
where the diffuson modes are introduced as
\begin{align*}
D_{\mu\nu}\left(i\omega,\mathbf{R}\right)\equiv\frac{1}{2}\left[\sigma_{\mu}\right]_{ji}D_{iji'j'}\left(i\omega,\mathbf{R}\right)\left[\sigma_{\nu}\right]_{i'j'}.
\end{align*}
Here $\sigma_0$ corresponds to the identity and the contraction in spin indices (latin labels) is assumed. Diagrammatically, $D_{iji'j'}$ corresponds to the boxes in the first diagram of Fig.~\ref{fig:diagrams}. At this point, it is useful to introduce the matrix
\begin{align*}
%\left[\hat{D}\left(i\omega,\mathbf{R}\right)\right]_{\mu\nu}\equiv D_{\mu\nu}\left(i\omega,\mathbf{R}\right).
\hat{D}\equiv \left(\begin{array}{cccc}
D_{00}&D_{0x}&D_{0y}&D_{0z}\\
D_{x0}&D_{xx}&D_{xy}&D_{xz}\\
D_{y0}&D_{yx}&D_{yy}&D_{yz}\\
D_{z0}&D_{zx}&D_{zy}&D_{zz}
\end{array}\right).
\end{align*}
In the diffusive regime, $\left|\omega\right|\tau\ll1$, with $\omega=\omega_1-\omega_2$ and $\text{sign}\left(\omega_1\right)\neq\text{sign}\left(\omega_2\right)$ (otherwise, the diffuson ladder is zero), the diffuson satisfy the equation
\begin{widetext}
\begin{align}
\left(\begin{array}{cccc}
\left|\omega\right|\tau-\tau D\nabla^2&0&0&0\\
0&\frac{1}{2}+\left|\omega\right|\tau-\tau D\nabla^2&0&\frac{1}{2}\tau v_F\partial_x\\
0&0&\frac{1}{2}+\left|\omega\right|\tau-\tau D\nabla^2&\frac{1}{2}\tau v_F\partial_y\\
0&-\frac{1}{2}\tau v_F\partial_x&-\frac{1}{2}\tau v_F\partial_y&1+\left|\omega\right|\tau-\tau D\nabla^2
\end{array}\right)\hat{D}\left(i\omega,\mathbf{r},\mathbf{r}'\right)=\delta^{(2)}\left(\mathbf{r}-\mathbf{r}'\right),
\label{eq:diffuson}
\end{align}
\end{widetext}
where $D=\frac{1}{2}v_F^2\tau$ is the diffusion constant. The triplet modes ($\mu,\nu=x,y,z$) are coupled by precession terms and only the singlet mode $D_{00}$ remains gapless %as consequence of the locking between spin and momentum implied by the Hamiltonian in Eq.~\eqref{eq:H_TI}
. Therefore, the triplet modes yields to an exponentially vanishing contribution to $\chi_{\alpha\beta}\chi_{\alpha'\beta'}\left(\mathbf{R}\right)$, similarly to what happens in a 2D electron gas with strong spin-orbit scattering \cite{Jagannathan}. Only the singlet mode contribution survives, leading in the $T\rightarrow0$ limit to
\begin{align}
\chi_{\alpha\beta}\chi_{\alpha'\beta'}\left(\mathbf{R}\right)=\frac{p_F^2}{24\pi^4 v_F^2R^4}\delta_{\alpha\alpha'}\delta_{\beta\beta'}.
\label{eq:2nd_moment}
\end{align}
As it is evident from this result, the RKKY coupling is still a long-range interaction, however, the variance of the interaction becomes independent of the orientation of the localized spins, $\left\langle\left(\text{H}_{\text{RKKY}}\right)^2\right\rangle_{dis}\sim\mathbf{S}_i^2\mathbf{S}_j^2$. %Due to the helicity of carriers, disorder fully randomizes the polarization induced by the presence of localized spins at distances larger than the mean free path.

Different magnetically ordered phases have been proposed in the literature \cite{Zhang1,Zhang2,Abanin_Pesin}. In diffusive media, however, these phases are expected to be suppressed due to the exponential decay of the first moment of the distribution for the exchange coupling, Eq.~\eqref{eq:1st_moment}, since, according to mean field arguments, it is this quantity what determines the critical temperature below which long-range magnetic order is possible. On the other hand, disorder introduces non-negligible fluctuations of the exchange coupling. In the particular case of an helical metal, the locking between momentum and spin of carriers makes the spin response independent of the orientation of the localized spins, Eq.~\eqref{eq:2nd_moment}. Due to these fluctuations, we expect that localized spins freeze to a non-zero value at low temperatures, but with no spatial correlation between them and zero net magnetization. In such a glassy phase, a global spectral gap is excluded.

We focus on the collective behavior of magnetic adatoms assuming that their spatial distribution is completely random. The clean limit with the Fermi level lying at the Dirac point was analyzed in Ref.~\onlinecite{Abanin_Pesin}. The form of the interaction in that case makes the in-plane interactions frustrated. Then, different behaviors are expected as a function of $\delta\equiv J_{\parallel}/J_{z}$. For $\delta\ll1$, a Ising-like ferromagnetic phase is expected at low temperatures, with magnetization along the out-of-plane axis. Hence, a gap in the spectrum of surface states is expected. In the opposite limit, $\delta\gg1$, the frustrated in-plane interactions dominate, giving rise to a spin glass separated from the Ising ferromagnet by a quantum critical point estimated to be located at $\delta_c\approx1.3$ \cite{Abanin_Pesin}. Similarly, fluctuations in the exchange interaction due to disorder tend to destroy long-range order, and the system freezes to a glass for sufficiently low temperatures.

We analyze in detail the situation when $\delta\ll1$. The spin Hamiltonian reads\begin{align*}
\text{H}_{\text{RKKY}}=-\sum_{i,j}J_{ij}\text{S}_i^z\text{S}_j^z,
\end{align*}
where the couplings $J_{ij}$ follow a certain statistical distribution. The first and second moments can be written in general as
\begin{gather*}
\left\langle J_{ij}\right\rangle_{dis}=J_{z}^2\sum_{\left\{\mathbf{R}\right\}}\mathcal{P}\left(\left\{\mathbf{R}\right\}\right)\chi_{zz}\left(\mathbf{R}\right),\nonumber\\
\left\langle \left(J_{ij}\right)^2\right\rangle_{dis}=J_{z}^4\sum_{\left\{\mathbf{R}\right\}}\mathcal{P}\left(\left\{\mathbf{R}\right\}\right)\chi_{zz}\chi_{zz}\left(\mathbf{R}\right).
\end{gather*}
Here $\mathcal{P}\left(\left\{\mathbf{R}\right\}\right)$ describes the statistical distribution of magnetic adatoms over the surface of the topological insulator. We assume for simplicity that they are uniformly distributed, so then we can approximate the above equations by \cite{foot_note} 
\begin{align}
\left\langle J_{ij} \right\rangle_{dis}\approx\frac{J_{z}^2}{\mathcal{A}}\int d^2\mathbf{R}\mbox{ }\chi_{zz}\left(\mathbf{R}\right)=
\frac{n_mJ_{z}^2p_F}{4v_FN}+\mathcal{O}\left(\frac{1}{p_F\ell}\right),
\label{eq:1st_moment_TI}
\end{align}
and similarly for the second moment,
\begin{gather}
\left\langle \left(J_{ij}\right)^2\right\rangle_{dis}\approx\frac{J_{z}^4}{\mathcal{A}}\int d^2\mathbf{R}\mbox{ }\chi_{zz}\chi_{zz}\left(\mathbf{R}\right)
\nonumber\\
=\frac{n_mJ_{z}^4p_F^2}{12\pi^3v_F^2N}\int_{\ell}^{\infty}dR\mbox{ }\frac{1}{R^3}=\frac{n_mJ_{z}^4p_F^2}{24\pi^3v_F^2\ell^2N}.
\label{eq:2nd_moment_TI}
\end{gather}
Here $\mathcal{A}$ is the area of the system and $N=n_m\mathcal{A}$, the number of localized spins. Note that the integral in Eq.~\eqref{eq:2nd_moment_TI} is infrared divergent, but the result of Eq.~\eqref{eq:2nd_moment} is strictly valid at $R\gg\ell$; therefore, there is a natural cut-off for this integral determined by the mean free path, which allows to estimate $\left\langle \left(J_{ij}\right)^2\right\rangle_{dis}$ to the leading order in $\left(p_F\ell\right)^{-1}$.

\begin{figure}
\begin{centering}
\includegraphics[width=1\columnwidth]{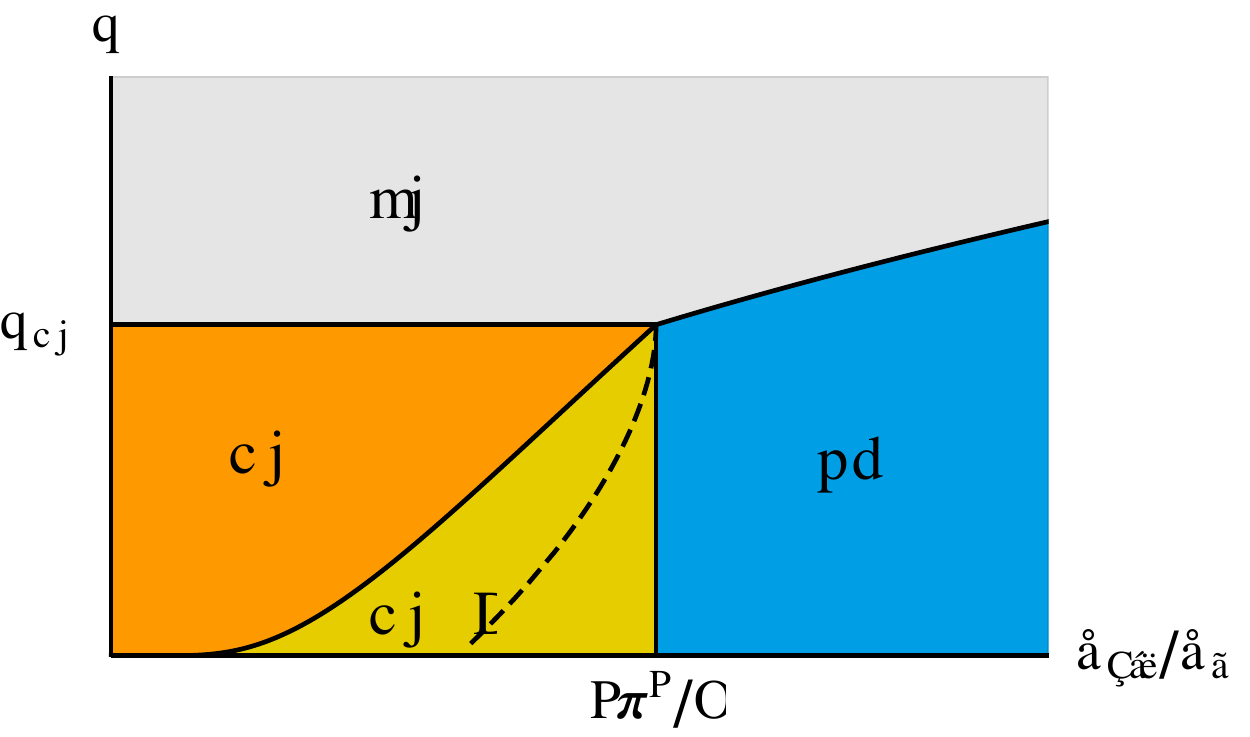}
\par\end{centering}
\caption{\label{fig:phase_diagram} Phase diagram of magnetic adatoms randomly distributed over the surface of a 3D topological insulator, showing paramagnetic (PM), ferromagnetic (FM), modified ferromagnetic (FM') and spin glass (SG) behavior. The results are based on the SK model for $\delta\ll 1$.}
\end{figure}

%Since in general $\left\langle \left(J_{ij}\right)^n\right\rangle=\mathcal{O}\left(\frac{1}{N}\right)$, 
We assume that higher moments of the distribution do not affect the thermodynamic properties of the system% \cite{Parisi_book}
, which can be studied within the Sherrington-Kirkpatrick (SK) model \cite{SK}. Following SK, the saddle-point equations reads\begin{gather}
\label{eq:m}
m=\int_{-\infty}^{\infty}\frac{dz}{\sqrt{2\pi}}e^{-\frac{1}{2}z^2}\tanh\left(\frac{T_{FM}}{T}m+\frac{T_{SG}}{T}q^{1/2}z\right),\\
q=\int_{-\infty}^{\infty}\frac{dz}{\sqrt{2\pi}}e^{-\frac{1}{2}z^2}\tanh^2\left(\frac{T_{FM}}{T}m+\frac{T_{SG}}{T}q^{1/2}z\right),
\label{eq:q}
\end{gather}
where $m=\left\langle\frac{1}{N}\sum_{i\in\text{m}}\left\langle \text{S}_i^z\right\rangle_T\right\rangle_{dis}$ is the magnetization per spin and $q=\left\langle\frac{1}{N}\sum_{i\in\text{m}}\left\langle \text{S}_i^z\right\rangle_T^2\right\rangle_{dis}$ is the Edwards-Anderson order parameter \cite{Edwards-Anderson}, and we have introduced \begin{gather*}
\label{eq:TFM}
T_{FM}=\frac{n_mJ_{z}^2p_F}{4v_Fk_B},\text{ and}\\
T_{SG}=\frac{J_{z}^2p_F}{2\pi v_Fk_B}\sqrt{\frac{n_mn_{dis}}{6\pi}}.
\label{eq:TSG}
\end{gather*}
Here we have taken $\ell\sim n_{dis}^{-1/2}$, where $n_{dis}$ represents the concentration of non-magnetic strong scatterers.

For low concentration of scatterers the model predicts a second order phase transition to a ferromagnetic state at $T=T_{FM}\propto n_m$. In the opposite limit, fluctuations in the couplings destroy long-range order, and a spin glass phase characterized by $q\neq0$ but $m=0$ is stabilized below $T_{SG}\propto\sqrt{n_mn_{dis}}$. The phase diagram is shown in Fig.~\ref{fig:phase_diagram}. Numerical solution of Eqs.~\eqref{eq:m}-\eqref{eq:q} predicts a region of the parameter space where the system passes from a paramagnetic to a ferromagnetic to a reentrant spin glass phase (dashed line in Fig.~\ref{fig:phase_diagram}) as the temperature is reduced. This is strange since at intermediate temperatures, where entropy plays a role, the system is ordered, but as $T\rightarrow0$ one finds that a disordered phase is preferred. This is associated to the instability of the saddle-point solution of the SK model and the replica symmetry breaking. From Parisi's solution \cite{Parisi} we predict a transition from the spin glass to a modified ferromagnetic phase (sometimes called mixed phase, where the replica symmetry is broken) at $n_{dis}=\frac{3\pi^3}{2}n_m$. The boundary between this modified ferromagnetic phase and a conventional Ising ferromagnet can be estimated from the Almeida-Thouless line \cite{Almeida-Thouless}.%,\begin{align*}
%\frac{T^2}{T_{SG}^2}=1-2q+\int\frac{dz}{\sqrt{2\pi}}e^{-\frac{1}{2}z^2}\tanh^4\left(\frac{T_{FM}}{T}m+\frac{T_{SG}}{T}q^{1/2}z\right).
%\end{align*}

Within this mean-field scenario, the formation of a spin glass at temperatures well below $T_{SG}$ can be seen as quenched fluctuations of a mass term in Eq.~\eqref{eq:H_TI}. Net magnetization is absent beyond the disorder threshold $n_{dis}=\frac{3\pi^3 n_m}{2}$, and the situation resembles the quantum percolation picture of the plateau transition of the quantum Hall effect \cite{Ludwig_etal}, where the mass plays the role of the control parameter. These fluctuations are marginally irrelevant in the renormalization group sense, whereas higher moments are marginal at the tree level, what justifies our previous assumptions.

In summary, we have computed the first moments of the distribution for the RKKY interaction in a weakly disorder helical metal and applied the results to infer some trends in the collective behavior of magnetic adatoms on a 3D topological insulator. For $J_{z}\gg J_{\parallel}$, our analysis reveals that the ensemble of localized spins freezes to a spin glass phase above the disorder threshold $n_{dis}=\frac{3\pi^3 n_m}{2}$. In-plane exchange interactions may help to stabilize this phase for weaker disorder due to frustration. The randomization in the orientation of the spins of the magnetic adatoms due to the helicity of carriers enforces this picture. In the spin glass phase, a spectral gap is precluded due to the lack  of a net magnetic moment. This means that an ARPES experiment would report no gap opening in this scenario. Therefore, the presence of disorder may explain the dispersion in results for photoemission experiments in magnetically-doped topological insulators. Nevertheless, a local probe \cite{sc_maps} could test the consequences of the breakdown of time reversal symmetry due to the freezing of the moments of the magnetic adatoms. 

Finally, it is worth to mention that the effect of a time-reversal symmetry breaking in the electron gas that mediates the interaction is not taken into account in a self-consistent way. In a recent calculation, the authors of Ref.~\onlinecite{Efimkin_Galitski} showed that a gap opening does not change the general structure of the RKKY interaction. Similarly, effects of dissipation can be neglected at first glance based on the argument that the RKKY interaction depends on electronic states deep inside the Fermi sea, not only at the Fermi surface, whereas dissipative phenomena as the Kondo effect is purely a Fermi-surface effect \cite{CastroNeto_Jones}. In other words, a rearrangement of the low energy part of the spectrum does not dramatically affect the RKKY interaction. Nevertheless, interesting novel physics beyond the scope of this work may appear at low temperatures, particularly in the disordered phase.

The author acknowledges support from the European Union's Seventh Framework Programme (FP7/2007-2013) through the ERC Advanced Grant NOVGRAPHENE (GA No. 290846), and useful discussions with A. Asenjo-Garcia, A. Cortijo and M. A. Cazalilla during the preparation of the manuscript.


\begin{thebibliography}{30}

\bibitem{rev} M. Z. Hasan and C. L. Kane, Rev. Mod. Phys. \textbf{82}, 3045 (2010); Xiao-Liang Qi and Shou-Cheng Zhang, \textit{ibid} \textbf{83}, 1057 (2011).

\bibitem{Qu_etal} Dong-Xia Qu, Y. S. Hor, Jun Xiong, R. J. Cava, and N. P. Ong, Science \textbf{329}, 821 (2010).

\bibitem{magnetic_films} Flavio S. Nogueira and Ilya Eremin, Phys. Rev. Lett. \textbf{109}, 237203 (2012); L\'aszl\'o Oroszl\'any and Alberto Cortijo, Phys. Rev. B \textbf{86}, 195427 (2012); S. V. Eremeev, V. N. Men'shov, V. V. Tugushev, P. M. Echenique, and E. V. Chulkov, \textit{ibid}. \textbf{88}, 144430 (2013).

\bibitem{Zhang1} Qin Liu, Chao-Xing Liu, Cenke Xu, Xiao-Liang Qi, and Shou-Cheng Zhang, Phys. Rev. Lett. \textbf{102}, 156603 (2009).

\bibitem{Zhang2} Jia-Ji Zhu, Dao-Xin Yao, Shou-Cheng Zhang, and Kai Chang, Phys. Rev. Lett. \textbf{106}, 097201 (2011).

\bibitem{Abanin_Pesin} D. A. Abanin and D. A. Pesin, Phys. Rev. Lett. \textbf{106}, 136802 (2011).

\bibitem{magneto-electric} Xiao-Liang Qi, Taylor L. Hughes, and Shou-Cheng Zhang, Phys. Rev. B \textbf{78}, 195424 (2008); Andrew M. Essin, Joel E. Moore, and David Vanderbilt, Phys. Rev. Lett. \textbf{102}, 146805 (2009); Ion Garate and M. Franz, \textit{ibid}. \textbf{104}, 146802 (2010).

\bibitem{general_yes} Y. L. Chen \textit{et al}., Science \textbf{329},  659 (2010); Y. S. Hor \textit{et al}., Phys. Rev. B \textbf{81} 195203 (2010); Su-Yang Xu \textit{et al}., Nat. Phys. \textbf{8}, 616 (2012); Joseph G. Checkelsky \textit{et al}., \textit{ibid}., 729 (2012).

\bibitem{general_no} T. Valla, Z.-H. Pan, D. Gardner, Y. S. Lee, and S. Chu, Phys. Rev. Lett. \textbf{108}, 117601 (2012); T. Schlenk \textit{et al}., \textit{ibid}. \textbf{110}, 126804 (2013); M. Ye \textit{et al}., Phys. Rev. B \textbf{85}, 205317 (2012); L. R. Shelford, T. Hesjedal, L. Collins-McIntyre, S. S. Dhesi, F. Maccherozzi, and G. van der Laan, \textit{ibid}. \textbf{86}, 081304(R) (2012).

\bibitem{nphys1838} L. Andrew Wray, Su-Yang Xu, Yuqi Xia, David Hsieh, Alexei V. Fedorov, Yew San Hor, Robert J. Cava, Arun Bansil, Hsin Lin, and M. Zahid Hasan, Nat. Phys. \textbf{7}, 32 (2011).

\bibitem{PRL108} M. R. Scholz, J. S\'anchez-Barriga, D. Marchenko, A. Varykhalov, A. Volykhov, L. V. Yashina, and O. Rader, Phys. Rev. Lett. \textbf{108}, 256810 (2012).

\bibitem{honolka2012} J. Honolka \textit{et al}., Phys. Rev. Lett. \textbf{108}, 256811 (2012).

%\bibitem{PRL107} Marco Bianchi, Richard C. Hatch, Jianli Mi, Bo Brummerstedt Iversen, and Philip Hofmann, Phys. Rev. Lett. \textbf{107}, 086802 (2011).

\bibitem{RKKY} M. A. Ruderman and C. Kittel, Phys. Rev. \textbf{96}, 99 (1954); T. Kasuya, Prog. Theor. Phys. \textbf{16}, 45 (1956); K. Yosida, Phys. Rev. \textbf{106}, 893 (1957).

\bibitem{deGennes} P. G. de Gennes, J. Phys. Radium \textbf{23}, 630 (1962).

\bibitem{rusos} A. Y. Zyuzin and B. Spivak, Pis'ma Zh. Eskp. Teor. Fiz. (JETP Lett.) \textbf{43}, 185 (234) (1986); L. N. Bulaevskii and S. V. Panyukov, \textit{ibid}., 190 (240) (1986).

\bibitem{Lee_Stone} P. A. Lee and A. D. Stone, Phys. Rev. Lett. \textbf{55}, 1622 (1985).

\bibitem{Jagannathan} A. Jagannathan, E. Abrahams, and M. J. Stephen, Phys. Rev. B \textbf{37}, 436 (1988); A. Jagannathan, Europhys. Lett. \textbf{17}, 437 (1992); Stefano Chesi and Daniel Loss, Phys. Rev. B \textbf{82}, 165303 (2010).

\bibitem{foot_note} Eq.~(2), which is strictly valid for $R\gg\ell$, reproduces the asymptotic result at $Rp_F\gg1$ for the RKKY interaction in the clean limit, $\ell\rightarrow\infty$. However, this must not be interpreted as an argument to extrapolate the result to any value of $\ell$. That becomes evident in the calculation of Eq.~(5), for which the result should be independent of the grade of disorder and proportional to $\gamma$. This is only true to the leading order in $\left(p_F\ell\right)^{-1}$, as the validity of Eq.~(2) ensures. This applies also for the calculation in Eq.~(6).

%\bibitem{Parisi_book} M. Mezard, G. Parisi, and M. A. Virasoro, \textit{Spin Glass Theory and Beyond}, (World Scientific, 1987).

\bibitem{SK} D. Sherrington and S. Kirkpatrick, Phys. Rev. Lett. \textbf{35}, 1792 (1975); S. Kirkpatrick and D. Sherrington, Phys. Rev. B \textbf{17}, 4384 (1978).

\bibitem{Edwards-Anderson} S. F. Edwards and P. W. Anderson, J. Phys. F: Metal Phys. \textbf{5}, 965 (1975).

\bibitem{Parisi} G. Parisi, J. Phys. A. \textbf{13}, 1101 (1980).

\bibitem{Almeida-Thouless} J. R. L. de Almeida and D. J. Thouless, J. Phys. A \textbf{11}, 983 (1978). %Note, however, that the differences between these two ferromagnetic phases are subtle since the Almeida-Thouless transition is third order in the Ehrenfest sense.

\bibitem{Ludwig_etal} Andreas W. W. Ludwig, Matthew P. A. Fisher, R. Shankar, and G. Grinstein, Phys. Rev. B \textbf{50}, 7526 (1994).

\bibitem{sc_maps} Paolo Sessi, Felix Reis, Thomas Bathon, Konstantin A. Kokh, Oleg E. Tereshchenko, and Matthias Bode, Nature Commun. \textbf{5}, 5349 (2014).

\bibitem{Efimkin_Galitski} D. K. Efimkin and V. Galitski, Phys. Rev. B \textbf{89}, 115431 (2014).

\bibitem{CastroNeto_Jones} A. H. Castro Neto and B. A. Jones, Phys. Rev. B \textbf{62}, 14975 (2000).

\end{thebibliography}
\end{document}